\documentclass[twocolumn,showpacs,preprintnumbers,amsmath,amssymb,a4paper,pre]{revtex4}
\usepackage{graphicx,color}
\usepackage{ulem}
\usepackage{amsmath}
\usepackage{amssymb}
\usepackage{verbatim}
\newcommand{\no}{\noindent}

\newcommand{\pdc}[1]{\ensuremath{\frac{\partial}{\partial #1}\,}}

\newcommand{\pddc}[1]{\ensuremath{\frac{\partial^2}{\partial #1 \,^2}\,}}
\begin{document}
\title{System size expansion for systems with an absorbing state}
\author{Francesca Di Patti}
\affiliation{Dipartimento di Fisica ``Galileo Galilei'',
Universit\`{a} degli Studi di Padova, via F. Marzolo 8, 35131
Padova, Italy }

\author{Sandro Azaele}
\affiliation{Institute of Integrative and Comparative Biology,
University of Leeds, Miall Building, Leeds LS2 9JT, United Kingdom}

\author{Jayanth R. Banavar}
\affiliation{Department of Physics, The Pennsylvania State
University, University Park, 104 Davey Laboratory, Pennsylvania
16802, USA}

\author{Amos Maritan}
\affiliation{Dipartimento di Fisica ``Galileo Galilei'',
Universit\`{a} degli Studi di Padova and INFN, via F. Marzolo 8, 35131
Padova, Italy}

\begin{abstract}
The well known van Kampen system  size expansion, while of rather
general applicability, is shown to fail to reproduce some
qualitative features of the time evolution for systems with an
absorbing state, apart from a transient initial time interval. We
generalize the van Kampen ansatz by introducing a new prescription
leading to non--Gaussian fluctuations around the absorbing state.
The two expansion predictions are explicitly compared for the
infinite range voter model with speciation as a paradigmatic model
with an absorbing state. The new expansion, both for a finite size
system in the large time limit and at finite time in the large size
limit, converges to to the exact solution as obtained in a numerical
implementation using the Gillespie algorithm. Furthermore, the
predicted lifetime distribution is shown to have the correct
asymptotic behavior.
\end{abstract}

\maketitle

The time evolution of systems consisting of large number of discrete
entities such as photons, nuclei, proteins or organisms is often
described by a master equation, a differential equation which, in
most cases, cannot be solved analytically. The van Kampen
system--size expansion \cite{gardiner,vanKampen} is one of the
techniques typically used to overcome such a limitation, although
alternative approaches have been proposed \cite{tome}. This method
allows one to account for the essential aspects of the problem and
provides a very useful tool to  approximate the temporal evolution.
However, such an approach is able to characterize the fluctuations
properly when the system has no boundaries or evolves far from them
\cite{alan}. For instance, if a system is driven towards ultimate
extinction, the van Kampen approximation is at best appropriate at
short times.

When a system with no boundaries initially has a large number of
particles, one expects that the macroscopic evolution is relatively
less affected by fluctuations at least within a finite temporal
scale. This general consideration leads to the rule of thumb that
deviations from the collective behavior are of order $\sqrt{N}$,
where $N$ is the size of the system. More specifically, the
population of the system, $n$, can be split into two contributions:
a macroscopic part of order $N$, $N\phi(t)$, whose evolution is
deterministic; and a random variable of order $\sqrt{N}$,
$\sqrt{N}\xi$. This is the celebrated van Kampen ansatz,
$n=N\phi(t)+\sqrt{N}\xi$, which approximates random jumps around the
macroscopic part with Gaussian fluctuations and naturally introduces
a small parameter for large $N$, $1/\sqrt{N}$, that can be used as
an expansion parameter for the solution of the master equation.
However, if the system has an absorbing state it will be driven
towards a final absorption, for example, eventual extinction. Thus,
sooner or later, the fluctuations could become comparable with the
macroscopic part, despite starting off with a large number of
individuals. This means that the validity of the van Kampen
approximation may be limited to a short initial time interval and
fluctuations may no longer be Gaussian.

As a paradigmatic example of a system with an absorbing state,
we will consider the infinite range voter model with
speciation, a simple model that can be handled analytically. By
exploiting standard methods used for diffusion processes with
absorbing boundaries, we will consider an improvement of the
classical van Kampen technique. However, we will show that
despite the modification, the new version fails to match
numerical simulations, thus calling for a different approach.
To provide a general context, we consider the following birth
and death master equation which is commonly encountered in
population ecology,
\begin{multline}\label{eq:masterEquation}
\frac{d}{dt}P_n(t) =  (\varepsilon_n^{-1} -1 )\left[ T(n+1|n) P_n(t)
\right ]  \\
+(\varepsilon_n^{+1} -1 )\left[T(n-1|n) P_{n}(t) \right]
\end{multline}
\no where $P_n(t)$ is the probability of observing the system in the
state $n$ at time $t$, and the shift operators $\varepsilon_n^{\pm
1}$ act on a function as $\varepsilon_n^{\pm 1}f(n) = f(n \pm 1)$.
$T(i|j)$ is the transition probability from state $j$ to state $i$.
In ecology, the state $n$ would correspond to a species abundance
$n$. When populations are large, customarily birth and death
transition rates turn out to be analytic functions of the density
$n/N$, namely $T(n\pm1|n)=T_{\pm}(n/N)$, where $N$ denotes the total
number of individuals or particles. This naturally suggests a
parameter which fluctuations can be compared to. Thus, in order to
obtain the correct system--size expansion we introduce the following
generalized van Kampen ansatz
\begin{equation}\label{eq:ansatz}
n= N \phi (t) + N^\alpha \xi
\end{equation}
\no with $0\leqslant \alpha < 1$. Under this assumption $n/N=\phi(t)
+ N^{\alpha-1}\xi$ and when $N\gg1$ the transition rates can be
expanded as power series of $y=\xi N^{\alpha-1}$ according to
\begin{equation}\label{eq:rates_expanded}
T_{\pm}(\phi+ y) = \sum_{k=0}^{\infty} T^{(k)}_{\pm}(\phi)
\frac{y^k}{k!}
\end{equation}
where $T^{(k)}_{\pm}(z)$ is the $k$--th derivative of $T_{\pm}(z)$.
Since $N$ is large the shift operators have the following
representation
\begin{equation}\label{eq:step}
\varepsilon_n^{\pm 1} \simeq 1 \pm  \frac{1}{N^{\alpha}}
\frac{\partial}{\partial \xi} + \frac{1}{2 N^{2\alpha}}
\frac{\partial^2}{\partial \xi^2 }\pm \ldots
\end{equation}

Substituting Eqs. (\ref{eq:ansatz}), (\ref{eq:rates_expanded}) and
(\ref{eq:step}) into Eq.(\ref{eq:masterEquation}) and defining the
new probability distribution $\Pi$ as $\Pi(\xi,t)\propto P_n(t)$,
one can collect terms proportional to different powers of $N$. In
the limit of large $N$ the leading order provides the usual
macroscopic law defined by the following deterministic equation
\begin{equation}\label{eq:macro_generica}
\frac{d}{d\tau} \phi = T_{+}(\phi)-T_{-}(\phi)
\end{equation}
\no where $\tau=t/N$ and we assume that
$T_{+}(\phi)-T_{-}(\phi)$ is not identically zero. Working out
the next--to--leading orders, one eventually obtains a
differential equation which up to the second derivative reads:
\begin{multline}\label{eq:equazione_con_rate_generici_alpha}
\frac{\partial}{\partial \tau} \Pi= N^0\Big [ T^{(1)}_{-}(\phi)-
T^{(1)}_{+}(\phi)\Big ] \frac{\partial}{\partial \xi} \Big ( \xi
\Pi\Big ) \\
+ \frac{1}{2} N^{1-2\alpha}\Big [ T_{-}(\phi)+ T_{+}(\phi)\Big ]
\frac{\partial^2}{\partial \xi^2} \Pi \\
+\frac{1}{2} N^{-\alpha}\Big [ T^{(1)}_{-}(\phi)+
T^{(1)}_{+}(\phi)\Big ]
\frac{\partial^2}{\partial \xi^2} \Big (  \xi \Pi \Big ) \\
+\sum_{k=2}^{\infty} \Big [ T^{(k)}_{-}(\phi)- T^{(k)}_{+}(\phi)\Big
] \frac{N^{(k-1)(\alpha-1)}}{k!}\frac{\partial}{\partial \xi} \Big (
\xi^k
\Pi \Big ) \\
+\sum_{k=2}^{\infty} \Big [
T^{(k)}_{-}(\phi)+T^{(k)}_{+}(\phi)\Big ] \frac{N^{k(\alpha-1)
-2 \alpha +1}}{k!}\frac{\partial^2}{\partial \xi^2} \Big (
\xi^k \Pi \Big )    ,
\end{multline}
where the time dependence of $\phi$ is given by Eq.
(\ref{eq:macro_generica}). The right--hand side of this
equation contains two series which are negligible with respect
to the first three terms when $N$ is large which are
proportional to $N^0$, $N^{1-2\alpha}$ and $N^{-\alpha}$
respectively. If we assume that both $ T^{(1)}_{-}(\phi)-
T^{(1)}_{+}(\phi)$ and $ T_{-}(\phi)+ T_{+}(\phi)$ are
different from zero in order to avoid the trivial result  of
vanishing fluctuations, one has to set $\alpha=1/2$ in
Eq.(\ref{eq:equazione_con_rate_generici_alpha}) Accordingly, in
the limit $N \rightarrow \infty$ we recover the standard van
Kampen equation
\begin{multline}\label{eq:gen_FP}
\frac{\partial}{\partial \tau} \Pi= \Big [ T^{(1)}_{-}(\phi)-
T^{(1)}_{+}(\phi)\Big ] \frac{\partial}{\partial \xi} \Big ( \xi
\Pi\Big ) \\
+ \frac{1}{2}\Big [ T_{-}(\phi)+ T_{+}(\phi)\Big ]
\frac{\partial^2}{\partial \xi^2} \Pi
\end{multline}
which is a linear Fokker--Planck equation whose solution is a
non--stationary Gaussian distribution.

However, for systems with absorbing boundaries at $n=0$ and large
temporal scales, the term proportional to $N^{1-2\alpha}$, $
T_{-}(\phi)+ T_{+}(\phi)$, approaches zero while the one
proportional to $N^{-\alpha}$, $T^{(1)}_{-}(\phi)+
T^{(1)}_{+}(\phi)$, does not. This is because when $\phi\rightarrow
0$, $ T_{-}(\phi) \pm T_{+}(\phi)$ are proportional to $\phi$ (in
the case of a simple absorbing state). In this case the van Kampen
prescription is no longer valid and we need to set $\alpha=0$ in the
limit of large $N$. For these systems fluctuations are progressively
more important in the long run, because $N\phi(\tau)\ll\xi$. Thus,
the differential equation governing the fluctuations is well
approximated by the following Fokker--Planck equation
\begin{multline}\label{eq:gen_sandro}
\frac{\partial}{\partial \tau} \Pi= \Big [ T^{(1)}_{-}(0)-
T^{(1)}_{+}(0)\Big ] \frac{\partial}{\partial \xi} \Big ( \xi
\Pi\Big ) \\
+\frac{1}{2}\Big [ T^{(1)}_{-}(0)+ T^{(1)}_{+}(0)\Big ]
\frac{\partial^2}{\partial \xi^2} \Big (  \xi \Pi \Big )
\end{multline}
This equation is different from Eq. (\ref{eq:gen_FP}) owing to
the linear diffusion term which results in the fluctuations
being no longer Gaussian distributed. Furthermore, if we do not
provide Eq. (\ref{eq:gen_FP}) with an absorbing boundary
condition for the solution, fluctuations could lead to negative
values for $n$. In contrast, Eq. (\ref{eq:gen_sandro}) has a
natural boundary at $\xi=0$ which prevents fluctuations, and
thus $n$, from becoming negative. Similar considerations also
hold when transition rates have more general algebraic
behaviors in the vicinity of the absorbing state
\footnote{Suppose that $f_{\pm}(x)$ are two analytic functions
such that $f_{\pm}(0)=1$, $\lambda_{\pm}$ two constants rates,
$l_{+}\geq l_{-}>1$ and that
$T(n\pm1|n)=\lambda_{\pm}(n/N)^{l_{\pm}}f_{\pm}(n/N)$ for
$n/N\ll1$. In this case  in Eq. (\ref{eq:gen_sandro}) both the
drift and diffusion terms are proportional to $\xi^{l_-}$.}.
Interestingly, Eq. (\ref{eq:gen_sandro}) can be exactly solved
\cite{lehnigk}, its solution being
\begin{multline}\label{eq:sol_sandro}
 \Pi(\xi,\tau|\xi_0,0)  =  \frac{\mu}{D}
  \frac{1}{1-e^{-\mu \tau}}\exp\left[-\frac{\frac{\mu}{D}(\xi+\xi_0e^{-
\mu \tau})}{1-e^{- \mu \tau}}\right] \\
\times \left(\frac{\xi}{\xi_0}e^{ \mu \tau}\right)^{-\frac{1}{2}}
I_{1}\left[\frac{\frac{2\mu}{D}\sqrt{\xi_0 \xi e^{\mu \tau}}}{e^{\mu
\tau}-1}\right]
\end{multline}
where $\mu= T^{(1)}_{-}(0)- T^{(1)}_{+}(0)$ is supposed to be
positive, $D=[T^{(1)}_{-}(0)+ T^{(1)}_{+}(0)]/2$, $I_{1}(z)$ is the
modified Bessel function of the first kind and $\xi_0$ is the value
of $\xi$ when $\tau=0$. It is worth noting that although the
solution is absorbing, one gets $\lim_{\xi\rightarrow
0}\Pi(\xi,\tau|\xi_0,0)\neq0$.
\begin{center}
\begin{figure}[tb]
\includegraphics[scale=0.3]{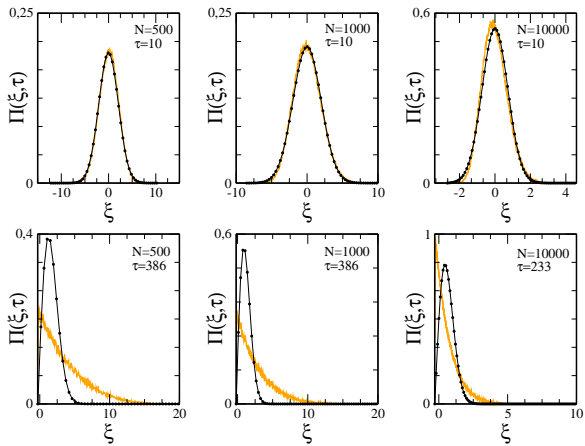}
\caption{Probability distribution for the fluctuations $\xi$ at
different times and with different parameters specified in the top
right of the panels. The noisy lines are the results of numerical solutions
obtained by averaging over $10^6$ stochastic realizations, the
solid--dotted lines represent the Gaussian solution modified for
absorbing boundaries given by (\ref{eq:vanK_abs}). Fluctuations are
approximately Gaussian distributed only for relatively short times,
while for large $\tau$ the Gaussian assumption breaks down. In all
cases $\nu=0.01$ and $n_0=300$.}\label{fig:greekPi}
\end{figure}
\end{center}

The theory we have developed so far can be applied straightforwardly
to many different absorbing systems. In particular, we now focus on
the infinite range voter model with speciation, a particular case of
the more general voter model which is of interest in opinion
formation problems \cite{liggett2004,redner2005,munoz2}, but also in
biological \cite{Evans1993,silvertown1992} and ecological contexts
\cite{durrett1994,zillio2005}.

The modified version of the voter model we investigate is
characterized by a parameter, the speciation rate $\nu$, which
averts the collapse of the whole system  into a trivial
monodominant state characterized by $\phi=1$. Specifically, let
us consider a system composed of $N$ elements, all of them
mutually interacting and belonging to possibly different
species. If we now focus on a specific species, we can re--map
all elements with two labels: the label $X_1$ for the elements
of the selected species, the label $X_0$ for the rest. Finally,
at each time step we randomly choose and update a pair of
elements according to the following interaction rules:
\begin{eqnarray}
X_1 + X_0  & \stackrel{1}{\longrightarrow} & X_0 + X_0 \label{eq:rule1}\\
X_0 + X_1  & \stackrel{1-\nu}{\longrightarrow} & X_1 + X_1 \label{eq:rule2}\\
X_1 + X_1  & \stackrel{\nu}{\longrightarrow} & X_1 + X_0   .
\label{eq:rule3}
\end{eqnarray}
An individual of the species of the first term on the lhs is
envisaged to be replaced by an individual of the second term on the
lhs except for speciation which occurs with a probability $\nu$ as
in the third rule. The factors above the arrow denote the
probability of the event indicated in the equation.

Let us denote by $n$ the number of $X_1$ individuals, so that $N-n$
is the total number of elements of type $X_0$. According to
(\ref{eq:rule1})--(\ref{eq:rule3}) the only transitions allowed are
those from $n$ to $n \pm 1$, and the corresponding transition
probabilities read
\begin{eqnarray}
T(n-1|n) & = & (1-\nu) \frac{n}{N} \frac{N-n}{N-1} + \nu
\frac{n}{N} \label{eq:T1}\\
T(n+1|n) & = & (1-\nu) \frac{N-n}{N}\frac{n}{N-1}  , \label{eq:T2}
\end{eqnarray}
where the initial states are on the right and final states on the
left. Since $T(\pm 1 | 0)=0$, once the population of $X_1$ dies out,
the selected species cannot be re--introduced into the system. Thus,
this model has a continual turn over  of species: new species appear
at rate $\nu$, but eventually they go extinct. This implies that
$n=0$ is an absorbing state. On the contrary, when the population of
$X_1$ reaches the maximum value $N$, transitions to $N+1$ are not
allowed since $T(N+1|N)= 0$, while $T(N-1|N)=\nu$. As a consequence,
$n=N$ is a reflecting boundary. In the following we will focus on
the time evolution of the system when $n$ is kept finite as $N$
becomes larger and larger \footnote{The master equation
(\ref{eq:masterEquation}) with the transition rates as given by Eqs.
(\ref{eq:T1}) and (\ref{eq:T2}) can be analytically solved in the
infinite size limit. However the explicit solution can be
numerically evaluated only at small $n$.}.

If we apply the generalized expansion described in the previous
section, we find that the macroscopic law according to Eq.
(\ref{eq:macro_generica}) is $\dot{\phi}=-\nu\phi$, thus
$\phi(\tau) = \phi_0 e^{-\nu \tau}$ with $\tau=t/(N-1)$. The
van Kampen equation corresponding to Eq. (\ref{eq:gen_FP})
reads
\begin{equation}\label{eq:FP}
\frac{\partial}{\partial \tau} \Pi= \nu \frac{\partial}{\partial
\xi} \left( \xi \Pi \right) + f(\tau) \frac{\partial^2}{\partial
\xi^2} \Pi
\end{equation}
where $f(\tau)=1/2\left[ (2-\nu)\phi_0 e^{-\nu \tau}+ 2(\nu-1)
\phi_0^2 e^{-2 \nu \tau} \right]$. Its solution is
\begin{equation*}
\Pi(\xi,\tau) = \frac{e^{\nu \tau}}{\sqrt{4 \pi \eta(\tau)}}
 \exp \left[ - \frac{(\xi e^{\nu \tau})^2 }{4
\eta(\tau)}\right]
\end{equation*}
with $ \eta(\tau)= (2-\nu) \phi_0 (e^{\nu \tau}-1)/(2
\nu)-(1-\nu)\phi_0^2 \tau$.

In order to account for the absorbing boundary, we added a time
dependent constraint on $\xi$, so that $n$ in Eq. (\ref{eq:ansatz})
varies between $0$ and $N$. To guarantee this latter condition, we
imposed  $\xi_{min} \leqslant \xi \leqslant\xi_{max}$, where
$\xi_{min}= -\sqrt{N} \phi_0 e^{-\nu \tau}$ and
$\xi_{max}=\sqrt{N}(1-\phi_0 e^{-\nu \tau})$. In correspondence to
$n=0$, $\xi=\xi_{min}$ is an absorbing boundary, while
$\xi=\xi_{max}$, which corresponds to $n=N$, is a reflecting
boundary. The final solution accounting for the absorbing boundary
reads
\begin{multline}\label{eq:vanK_abs}
\Pi_{abs}(\xi,\tau) = \frac{e^{\nu \tau}}{\sqrt{4 \pi \eta(\tau)}}
\left \{ \exp \left[ - \frac{(\xi e^{\nu \tau})^2 }{4
\eta(\tau)}\right] \right .\\
\left . -  \exp \left[ - \frac{(\xi e^{\nu \tau} + 2 \sqrt{N}
\phi_0)^2 }{4 \eta(\tau)}\right] \right \}
\end{multline}
This solution has a delta peak at $\xi=0$ (or $n=N\phi_0$) for
$\tau\rightarrow 0$ and vanishes at $\xi=-\sqrt{N} \phi_0 e^{-\nu
\tau}$.

We now turn to Eq. (\ref{eq:gen_sandro}) which now reads
\begin{equation*}
\frac{\partial}{\partial \tau}\Pi(\xi,\tau)=\nu
\pdc{\xi}\left(\xi\Pi\right)+
\frac{2-\nu}{2}\pddc{\xi}\left(\xi\Pi\right)
\end{equation*}
where we have used that  $T^{(1)}_{-}(0)- T^{(1)}_{+}(0)=\nu$ and
$T^{(1)}_{-}(0)+ T^{(1)}_{+}(0)=2-\nu$ (see Eqs. (\ref{eq:T1}) and
(\ref{eq:T2})).

\begin{center}
\begin{figure}[tb]
\includegraphics[scale=0.3]{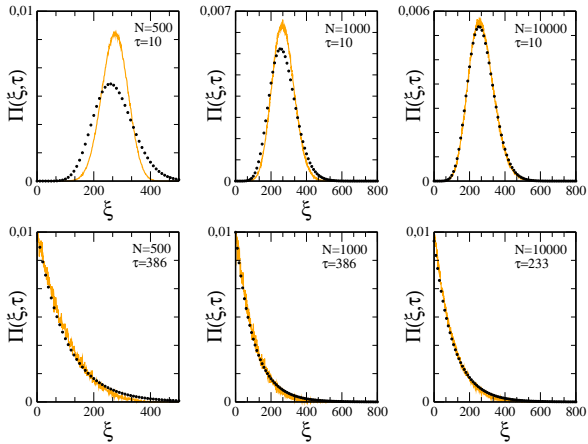}
\caption{Comparison between the numerical simulations of the rules
in Eqs. (\ref{eq:rule1})--(\ref{eq:rule3}) and the theoretical
predictions of the probability distribution in Eq.
(\ref{eq:sol_sandro}) at different times and systems sizes,
according to the parameters specified in the top right of the
panels. The noisy lines are the numerical profiles obtained by
averaging over $10^6$ stochastic simulations. While initially
fluctuations are approximately Gaussian, in the sequel they are
non--Gaussian and peak at zero. In all cases $\nu=0.01$ and
$n_0=300$.}\label{fig:mean_field}
\end{figure}
\end{center}

In order to test the validity of the methods, we performed
extensive numerical simulations of  Eqs.
(\ref{eq:rule1})--(\ref{eq:rule3}) through the Gillespie
algorithm \cite{gillespie}, which allows one to produce time
series which exactly recover the solution of the master
equation (\ref{eq:masterEquation}) with the rates in Eqs.
(\ref{eq:T1}) and (\ref{eq:T2}). Fig. (\ref{fig:greekPi}) shows
typical results of the stochastic simulations and their
comparison with the absorbing van Kampen solution in Eq.
(\ref{eq:vanK_abs}). For short times the first three profiles
overlap well, but as time increases, the probability
distribution does not match the numerical simulation, as shown
in the last three panels. Furthermore, the agreement does not
improve on increasing the size of the system. In contrast, Fig.
(\ref{fig:mean_field}) shows the solution in Eq.
(\ref{eq:sol_sandro}) with $\mu=\nu$ and $D=(2-\nu)/2$.
Increasing $N$, while keeping  $\tau$ fixed, improves the
matching as shown in the first three panels. As expected Eq.
(\ref{eq:sol_sandro}) converges to the numerical profiles as
$\tau$ increases.

As illustrated in both figures, in the presence of an absorbing
state, the system is characterized by at least two temporal
scales, $\tau_{1}$ and $\tau_{2}$, which make fluctuations
evolve according to Eq. (\ref{eq:gen_FP}) for $\tau < \tau_{1}$
and Eq. (\ref{eq:gen_sandro}) for $\tau > \tau_{2}$. It is
possible to estimate roughly the two scales by observing that
one should expect the generalized van Kampen ansatz to work
until the fluctuations are of the same magnitude as the
macroscopic part, namely $N\phi(\tau) \simeq N^{\alpha}
\sigma(\tau)$ where $\sigma(\tau)$ is the variance of $\xi$.
When $\alpha=1/2$, this condition translates into $ N\phi_0
e^{-\nu \tau} \simeq \sqrt{2N\eta(\tau)}$ which gives $\tau_{1}
\simeq 1/(3\nu) \ln(n_0\nu/(2-\nu))$ for $\nu \tau\gg1$. For
the expansion with $\alpha=0$, we have $ N \phi_0 e^{-\nu \tau}
\simeq e^{-\nu \tau} \sqrt{(2-\nu)n_0/\nu (e^{\nu \tau}-1) }$
which gives $\tau_{2} \simeq 1/\nu \ln(\nu n_0/(2-\nu)+1)$.
Note that $\tau_{1}<\tau_{2}$. The above mentioned condition of
validity of the classic van Kampen expansion is confirmed by
numerical simulations (data not shown).

Finally, for systems with absorbing boundaries, it is interesting to
calculate an analytical expression for the survival probability
$P_{S}(\tau)$ \cite{munoz1}. In our case, we get the exact
expression $P_{S}(\tau)=1-\left [ (1-e^{-\nu
\tau})/(1-(1-\nu)e^{-\nu \tau}) \right ]^{n_0}$ which, in the
scaling limit $\nu \rightarrow 0$ with $\nu \tau$ fixed, simplifies
to $f(\nu \tau)/t$ with $f(z)=zn_0/(e^{z}-1)$. Amazingly, the same
result is also obtainable using Eq. (\ref{eq:sol_sandro}), with
$P_{S}(\tau)=\int d \xi \Pi(\xi, \tau | \xi_0,0)$.

Summarizing, in the presence of systems with absorbing states,
one has to generalize the standard van Kampen ansatz in order
to monitor the temporal evolution at large times. As time
elapses, fluctuations become more and more important and are no
longer Gaussian. However, they still can be  analytically
treated and lead to the general solution given by Eq.
(\ref{eq:sol_sandro}).

Acknowledgments. We thank Duccio Fanelli for useful discussions. S.
A. acknowledges  the EU FP7 SCALES project (No. 26852) for financial
support. The work is supported by The Cariparo foundation.

\bibliographystyle{apsrev4-1}
\bibliography{bibliography}
\end{document}